\documentclass[twocolumn,prx,preprintnumbers,showpacs,amsmath,amssymb,superscriptaddress]{revtex4-1}

\pdfoutput=1

\usepackage[pdftex, colorlinks, citecolor=blue]{hyperref}   
\usepackage{graphicx}
\usepackage{dcolumn}
\usepackage{txfonts}
\usepackage{xcolor}
\usepackage{amssymb}
\usepackage{amsmath}
\usepackage{latexsym}
\usepackage{epsfig}
\usepackage{amsbsy}
\usepackage{array}
\usepackage{tabularx}

\usepackage[cp1252]{inputenc}
\usepackage{tikz-feynman}
\usepackage{comment}
\usepackage{algcompatible}
\usepackage{float}

\newenvironment{alg}%
{\vskip10pt\noindent{\rule{0.45\textwidth}{.4pt}}\par\nobreak\vskip-5pt}%
{\vskip-5pt\noindent{\rule{0.45\textwidth}{.4pt}}\par\nobreak\vskip0pt}%

\newcommand{\bra}[1]{\langle #1 \mid}
\newcommand{\ket}[1]{\mid #1 \rangle}

\newcommand{\sais}{\ifmmode\mbox{\c{S}}\else\c{S}\fi{}a\ifmmode\mbox{\c{s}}\else\c{s}\fi{}\ifmmode
\imath \else \i \fi{}o\ifmmode \breve{g}\else \u{g}\fi{}lu~}

\begin{document}

\title{Diagrammatic Monte Carlo study of the Fr\"ohlich polaron dispersion in 2D and 3D}

\author{Thomas Hahn}
\affiliation{University of Vienna, Faculty of Physics and Center for
Computational Materials Science, Sensengasse 8, A-1090 Vienna, Austria}

\author{Sergei Klimin}   
\affiliation{Theory of Quantum and Complex Systems, Universiteit Antwerpen, Universiteitsplein 1, B-2610 Antwerpen, Belgium}

\author{Jacques Tempere}
\affiliation{Theory of Quantum and Complex Systems, Universiteit Antwerpen, Universiteitsplein 1, B-2610 Antwerpen, Belgium}

\author{Jozef T. Devreese}
\affiliation{Theory of Quantum and Complex Systems, Universiteit Antwerpen, Universiteitsplein 1, B-2610 Antwerpen, Belgium}

\author{Cesare Franchini}
\affiliation{University of Vienna, Faculty of Physics and Center for
Computational Materials Science, Sensengasse 8, A-1090 Vienna, Austria}

\begin{abstract}
We present results for the solution of the large polaron Fr\"ohlich Hamiltonian in 3-dimensions (3D) and 2-dimensions (2D) obtained via the Diagrammatic Monte Carlo (DMC) method. Our implementation is based on the approach by Mishchenko [A.S. Mishchenko \textit{et al.}, Phys. Rev. B \textbf{62}, 6317 (2000)]. Polaron ground state energies and effective polaron masses are successfully benchmarked with data obtained using Feynman's path integral formalism. By comparing 3D and 2D data, we verify the analytically exact scaling relations
for energies and effective masses from 3D$\to$2D, which provides a stringent test for the quality of DMC predictions. The accuracy of our results is further proven by providing values for the exactly known coefficients in weak- and strong coupling expansions. 
Moreover, we compute polaron dispersion curves which are validated with analytically known lower and 
upper limits in the small coupling regime and verify the first order expansion results for larger couplings, thus disproving previous critiques on the apparent incompatibility of DMC with analytical results and furnishing 
useful reference for a wide range of coupling strengths.
\end{abstract}

\maketitle

\section{Introduction} \label{sec:introduction}

Ever since the emergence of polaron theory in the 1930s~\cite{Landau1933}, the concept of polarons has been applied to a wide variety of physical systems in which a particle is coupled to its environment, e.g. spin or magnetic polarons~\cite{PSSB:PSSB2220650102}, exciton polarons~\cite{Haken}, BEC-impurity polarons~\cite{PhysRev.127.1452}, ripplonic polaron~\cite{PhysRevB.24.499} etc. The polaron problem in its original form considers a single electron in a polar crystal interacting with the surrounding lattice. Due to Coulomb forces, the electron distorts the ions in its neighbourhood, which creates a polarization that follows the electron as it moves through the crystal. This generated polarization acts back on the electron and so renormalizes electronic properties. The resulting quasiparticle consisting of the electron surrounded by the distorted lattice was termed a "polaron". Nowadays (cf. the review by Alexandrov and Devreese~\cite{JTDbook}) a more quantum mechanical picture of a polaron is used in which the electron dresses itself with a cloud of phonons. 

Polarons may be classified according to the strength of the electron-phonon coupling (weak/strong) and the extension of the lattice distortion around the electron (small/large)~\cite{JTDbook,Rashba2005347}. Weak-coupling polarons dress themselves with only a small number of phonons $\bar{N} \ll 1$ leading to a slightly enhanced effective mass compared to the "bare" electron $(m_* - m) \ll m$. Strong-coupling polarons have more phonons in the cloud $\bar{N} \gg 1$ and a much larger effective mass $m_*/m \gg 1$. By $\bar{N}$ we denote the average number of phonons in the cloud, $m_*$ is the effective mass of the polaron and $m$ the mass of the "bare" electron without coupling. Furthermore, a polaron is called a small polaron when the lattice distortion induced by the electron is of the same size as the lattice constant and a large polaron when the distortion extends over several lattice sites. 
Typically, the description of small polarons requires the treatment of short-range electron-phonon interaction and an explicit account of the lattice periodicity. Instead, the theory of large polarons assumes long-range forces and relies on the continuum approximation.

Studies of polarons are historically conducted in the framework of quantum field theory using effective quantum Hamiltonians~\cite{Frohlich1954,holstein1959ann}. 
More recently, first principles methods based on density functional theory turned out to provide an accurate microscopic description of both large and small polarons~\cite{Setvin,Miyatae1701217}.
The most famous model Hamiltonians go back to the 1950s to Fr\"ohlich~\cite{Frohlich1954} and Holstein~\cite{holstein1959ann}. Both contain a term for a free particle $H_\textrm{e}$, a free phonon field $H_{\textrm{ph}}$ and for the particle-phonon interaction $H_{\textrm{e-ph}}$. While the Holstein Hamiltonian models small polarons, the Fr\"ohlich Hamiltonian, which is the focus of the present study, describes large polarons and is given as 
\begin{gather}
H = H_\textrm{e} + H_{\textrm{ph}} + H_{\textrm{e-ph}}, \label{eqn:fullHamiltonian} \\
H_\textrm{e} = \sum_\mathbf{k} \frac{k^2}{2} a_\mathbf{k}^{\dagger}a_\mathbf{k}^{\vphantom{\dagger}}, \label{eqn:electronHamiltonian} \\
H_{\textrm{ph}} = \sum_\mathbf{q}b_\mathbf{q}^{\dagger}b_\mathbf{q}^{\vphantom{\dagger}}, \label{eqn:phononHamiltonian} \\
H_{\textrm{e-ph}} = \sum_{\mathbf{k,q}}\left[V_{d}^{\vphantom{\dagger}}(\mathbf{q})b_\mathbf{q}^{\dagger}a_\mathbf{k-q}^{\dagger}a_\mathbf{k}^{\vphantom{\dagger}} + V^{\dagger}_{d}(\mathbf{q})b_\mathbf{q}^{\vphantom{\dagger}}a_\mathbf{k+q}^{\dagger}a_\mathbf{q}^{\vphantom{\dagger}}\right]. \label{eqn:interactionHamiltonian}
\end{gather}
Here $a_\mathbf{k}^{\vphantom{\dagger}}$ and $b_\mathbf{q}^{\vphantom{\dagger}}$ are destruction operators for a particle with wave vector $\mathbf{k}$ and a phonon with wave vector $\mathbf{q}$, respectively. $V_{d}^{\vphantom{\dagger}}(\mathbf{q})$ is the coupling function for a system in $d$ dimensions and takes the form
\begin{equation}
V_3^{\vphantom{\dagger}}(\mathbf{q}) = i \left(\frac{2\sqrt{2}\pi\alpha}{A}\right)^{\frac{1}{2}}\frac{1}{q} \label{eqn:coupling3d}
\end{equation}
in 3 dimensions and
\begin{equation}
V_2^{\vphantom{\dagger}}(\mathbf{q}) = i \left(\frac{\sqrt{2}\pi\alpha}{A}\right)^{\frac{1}{2}}\frac{1}{\sqrt{q}} \label{eqn:coupling2d}
\end{equation}
in 2 dimensions~\cite{PhysRevB.33.3926}. In Eq.~\ref{eqn:coupling3d} and~\ref{eqn:coupling2d}, $A$ is the $d$-dimensional volume of the system and $\alpha$ is the coupling constant which is material dependent and determines the strength of the electron-phonon interaction. Typical values for real materials are in the range $0<\alpha<5$~\cite{1402-4896-1989-T25-056}. Units are chosen such that energy is measured in units of $\hbar\omega_0$ and length in units of $\sqrt{\hbar/m\omega_0}$ which leads to $\hbar=\omega_0=m=1$. In deriving and solving the Fr\"ohlich Hamiltonian, it is a common practice to assume certain approximations: (i) the energy dispersion for the electron is parabolic with a band mass $m$, (ii) the phonon frequency $\omega(\mathbf{q})=\omega_0$ is dispersionless and constant, (iii) the interaction is only between the electron and long-wavelength optical, longitudinal phonons and (iv) the spatial extension of the polaron is larger than the lattice constant. In this paper, we exclusively focus on the Fr\"ohlich model 
 and we study the polaron dispersion law, i.e. the dependence of the ground-state energy $E_0(k,\alpha)$ on the modulus of the total polaron momentum $k=|\textbf{k}|$. 

A large body of work~\cite{JTDbook} exists on solving the Fr\"ohlich Hamiltonian, and most of it 
concerns the energy of the polaron at rest, $E_0(0,\alpha)$. Yet, so far no exact analytic solution 
was found. The most successful approach to calculate $E_0(0,\alpha)$ is Feynman's path integral
formalism~\cite{Feynman,Rosenfelder2001}, a variational treatment that provides a very 
accurate upper bound for 
the polaron ground state energy for all coupling strengths as well as approximate values for 
the polaron effective mass. Early work on the behavior of the dispersion
curve~\cite{Whitfield1965,Appel1968} allowed to conclude that the
energy-momentum relation starts off quadratically at low $k$ (thus allowing to define a polaron mass)
but bends over when approaching the continuum edge $E_c(\alpha)=E_0(0,\alpha)+\hbar \omega_0$. Later  
it was found that in 3D the dispersion hits the continuum edge whereas for 2D it 
approaches it asymptotically, and upper and lower bounds for the dispersion were 
obtained~\cite{PhysRevB.60.10886,Gerlach2003,PhysRevB.77.174303}. These bounds, as well as some
analytically known limits, constitute good benchmarks for any theory of the polaron dispersion.

More recently, the Diagrammatic Monte Carlo method (DMC) was developed and applied to the 3-dimensional Fr\"ohlich polaron~\cite{Prokofev1998,Mishchenko2000}. It makes use of diagrammatic expansions of Green's functions and a Metropolis sampling algorithm to perform a random walk in the space of all Feynman diagrams. The DMC not only allows for the calculation of the ground state energies but as well as the polaron dispersion curves, Z-factors (quasiparticle weights) and phonon statistics.
However, the DMC results~\cite{Prokofev1998,Mishchenko2000} were 
criticized~\cite{PhysRevB.77.174303,Gerlach2003}: the reported results disagree with 
the analytically known second order coefficient in $\alpha$ for the polaron ground state energy, 
as well as the large-$\alpha$ expansion coefficient.

The aim of the present paper is the application of our newly implemented DMC code to the solution 
of the Fr\"ohlich Hamiltonian in both the 3-dimensional (3D) and the 2-dimensional (2D) case.
To our knowledge, there do not exist any DMC results for the 2D Fr\"ohlich polaron in the literature.
We find that the present DMC results, both in 2D and 3D, agree with the analytically 
known limits, thus refuting the critique of the DMC method formulated 
in~\cite{PhysRevB.77.174303,Gerlach2003}. In addition, we compare the obtained dispersion relations with
analytic upper and lower bounds (where available) and a fitting function~\cite{PhysRevB.77.174303}.

The structure of the paper is as follows. 
The DMC program is based on the seminal works of Prokof'ev ~\cite{Prokofev1998} and Mishchenko~\cite{Mishchenko2000}, and is described in Sec.~\ref{sec:theory}. The numerical outcome is presented and discussed in
Sec.~\ref{sec:results}. We first benchmark our results for the 3D case with the reference data of Prokof'ev \textit{et al.}~\cite{Prokofev1998} and Mishchenko \textit{et al.}~\cite{Mishchenko2000} as well as with results obtained from Feynman's path integral approach~\cite{Rosenfelder2001}. Furthermore, we show ground state energies $E_{0}(0,\alpha)$, polaron dispersions $E_{0}(k,\alpha)$ and effective masses $m_{*}(\alpha)$ for the 2D Fr\"ohlich polaron and compare them to various scaling relations derived by Peeters and Devreese~\cite{PhysRevB.36.4442}. We also provide values for the exactly known weak- and strong coupling coefficients.  Finally, conclusive remarks are drawn in Sec.~\ref{sec:conclusion}.

\section{Theory and Methodology} \label{sec:theory}

In this section, we introduce the concepts of many-body Green's functions, diagrammatic expansions and corresponding Feynman diagrams as well as the basic concepts of the Diagrammatic Monte Carlo method. Necessary computational details of our code are also given in this section.

\subsection{Green's functions and Feynman diagrams} \label{subsec:green}

To solve the Fr\"ohlich Hamiltonian from Eq.~\ref{eqn:fullHamiltonian} for the lowest energy eigenvalues, we make use of the Green's function formalism from many-body physics. In particular, we are interested in the one-electron-$N$-phonon Green's function in the momentum ($\mathbf{k},\tilde{\mathbf{q}}_i$) - imaginary time ($\tau$) representation at zero-temperature, where we assume $\tau>0$:
\begin{gather}
\begin{split}
G^{(N)}(\mathbf{k},\tau,\{\tilde{\mathbf{q}}_i\}) = & \bra{0}b_{\tilde{\mathbf{q}}_N}^{\vphantom{\dagger}}(\tau)\dots b_{\tilde{\mathbf{q}}_1}^{\vphantom{\dagger}}(\tau)a_{\mathbf{k}_1}^{\vphantom{\dagger}}(\tau) \\
& a_{\mathbf{k}_1}^{\dagger}(0)b_{\tilde{\mathbf{q}}_1}^{\dagger}(0)\dots b_{\tilde{\mathbf{q}}_N}^{\dagger}(0)\ket{0}.
\end{split} \label{eqn:Ngreen}
\end{gather}
The ket $\ket{0}$ in Eq.~\ref{eqn:Ngreen} is the electron and phonon vacuum state~\cite{Pines} and the operators are in the Heisenberg picture $a_\mathbf{k}^{\vphantom{\dagger}}(\tau)=e^{\tau H}a_\mathbf{k}^{\vphantom{\dagger}}e^{-\tau H}$. The total or polaron wave vector is given by $\mathbf{k}=\mathbf{k}_1+\sum_\mathbf{i}\tilde{\mathbf{q}}_i$ and is a conserved quantity~\cite{Frohlich1954}.

By adding a complete set of polaron eigenstates $\ket{\beta(\mathbf{k})}$ to Eq.~\ref{eqn:Ngreen}, with $H\ket{\beta(\mathbf{k})}=E_{\beta}(\mathbf{k})\ket{\beta(\mathbf{k})}$ and $H\ket{0}=E_v\ket{0}=0$, the Green's function becomes
\begin{eqnarray}
G^{(N)}(\mathbf{k},\tau,\{\tilde{\mathbf{q}}_i\}) & = & \sum_\beta \big|\bra{\beta(\mathbf{k})}a_{\mathbf{k}_1}^\dagger b_{\tilde{\mathbf{q}}_1}^\dagger \dots b_{\tilde{\mathbf{q}}_N}^\dagger \ket{0}\big|^2 e^{-(E_\beta(\mathbf{k})-E_v)\tau}= \nonumber \\
& = & \sum_\beta Z_\beta^{(N)}\left(\mathbf{k},\{\tilde{\mathbf{q}}_i\}\right) e^{-E_\beta(\mathbf{k})\tau}. \label{eqn:NgreenExpanded}
\end{eqnarray}
The $Z_\beta^{(N)}$-factor measures the squared overlap between the polaron eigenstate $\ket{\beta(\mathbf{k})}$ and a state with one free electron and $N$ free phonons. If $\tau \to \infty$, Eq.~\ref{eqn:NgreenExpanded} shows that the term which contains the state with the lowest energy eigenvalue $E_{0}(\mathbf{k})$ is the dominant one in the sum. Therefore it is possible to retrieve $E_{0}(\mathbf{k})$ and the corresponding $Z_{0}^{(N)}\left(\mathbf{k},\{\tilde{\mathbf{q}}_i\}\right)$-factor for given $\mathbf{k}$ and $\{\tilde{\mathbf{q}}_i\}$ values from the asymptotic behaviour of the Green's function at long imaginary-times:
\begin{equation}
G^{(N)}(\mathbf{k},\tau\to\infty,\{\tilde{\mathbf{q}}_i\})=Z_{0}^{(N)}\left(\mathbf{k},\{\tilde{\mathbf{q}}_i\}\right)e^{-E_{0}(\mathbf{k})\tau}. \label{eqn:greenAsympt}
\end{equation}

To calculate $G^{(N)}$, we expand the Green's function in a perturbation series~\cite{mahan2000many}. Formally, this leads to an expression of the form 
\begin{equation}
G^{(N)}(\mathbf{k},\tau,\{\tilde{\mathbf{q}}_i\}) = \sum_{n=0}^\infty \sum_{\xi_n} \idotsint \mathcal{D}_{n,\xi_n}\big(\mathbf{k},\tau,\{\tilde{\mathbf{q}}_i\};\mathbf{x}\big) \ d\mathbf{x}, \label{eqn:greenPertExpansion}
\end{equation}
where $n$ labels the order of the perturbation expansion, $\xi_n$ indexes different terms of the same order and $\mathbf{x}=(\tau_1,\dots,\tau_n,\mathbf{q}_1,\dots,\mathbf{q}_{k})$ is a vector of integration variables (times of interaction vertices and internal phonon wave vectors). Note the difference between external phonon wave vectors $\{\tilde{\mathbf{q}}_i\}$ appearing in the definition of $G^{(N)}$ and internal phonon wave vectors $\{\mathbf{q}_i\}$ over which is integrated. The integrands $\mathcal{D}_{n,\xi_n}$ are given as a product of free electron Green's functions $G_0(\mathbf{k},\tau_i-\tau_j)$, free phonon Green's functions $W_0(\mathbf{q},\tau_i-\tau_j)$ and squared interaction vertices $|V_d^{\vphantom{\dagger}}(\mathbf{q})|^2$. With the following simple rules it is possible to map all $\mathcal{D}_{n,\xi_n}$ functions to Feynman diagrams:
\begin{gather}
G_0(\mathbf{k},\tau_i-\tau_j)=
\begin{tikzpicture}[baseline=-3pt]
\begin{feynman}
\vertex [label=below:{$\tau_j$}](a);
\vertex [anchor=base,right=2cm of a,label=below:{$\tau_i$}] (b);
\diagram* {
(a) -- [fermion, edge label'=\footnotesize \(\mathbf{k}\)] (b),
};
\end{feynman}
\end{tikzpicture}
= e^{-k^2/2(\tau_i-\tau_j)}, \label{eqn:freeElectron} \\
W_0(\mathbf{q},\tau_i-\tau_j)=
\begin{tikzpicture}[baseline=-3pt]
\begin{feynman}
\vertex [label={[shift={(0.1,-0.65)}]$\tau_j$}](a) {};
\vertex [anchor=base,right=2cm of a,label={[shift={(-0.1,-0.65)}]$\tau_i$}] (b) {};
\diagram* {
(a) -- [boson, edge label'=\footnotesize \(\mathbf{q}\)] (b),
};
\end{feynman}
\end{tikzpicture}
= e^{-\omega_0(\tau_i-\tau_j)}, \label{eqn:freePhonon} \\
|V_d^{\vphantom{\dagger}}(\mathbf{q})|^2 =
\begin{tikzpicture}[baseline=-2pt]
\begin{feynman}
\vertex (a);
\vertex [right=0.5cm of a,dot,label=below:{\footnotesize $V_d^{\vphantom{\dagger}}(\mathbf{q})$}] (b) {};
\vertex [right=1.5cm of b,dot,label=below:{\footnotesize $V_d^\dagger(\mathbf{q})$}] (c) {};
\vertex [right=0.5cm of c] (d) {};
\vertex [above=0.75cm of b] (e) {};
\vertex [above=0.75cm of c] (f) {};
\diagram* {
(a) -- (b) -- [fermion] (c) -- (d),
(b) -- [boson, edge label=\footnotesize \(\mathbf{q}\)] (e),
(c) -- [boson, edge label=\footnotesize \(\mathbf{q}\)] (f),
};
\end{feynman}
\end{tikzpicture}
=\frac{(d-1)\sqrt{2}\pi\alpha}{Aq^{d-1}}. \label{eqn:vertex}
\end{gather}

\begin{figure}
\centering
\begin{tikzpicture}
\begin{feynman}
\vertex (a) [label=below:{$0$}];
\vertex [right=7cm of a,label={[shift={(-0.1,-0.55)}]$\tau$}] (b) {};
\vertex [above=0.7cm of a] (c) {};
\vertex [above=3cm of b] (d) {};
\vertex [above=2cm of a] (e) {};
\vertex [above=1.5cm of b] (f) {};
\vertex [right=0.7cm of a,dot,label=below:{$\tau_1$}] (v1) {};
\vertex [right=1.3cm of a,dot,label=below:{$\tau_2$}] (v2) {};
\vertex [right=2cm of a,dot,label=below:{$\tau_3$}] (v3) {};
\vertex [right=3cm of a,dot,label=below:{$\tau_4$}] (v4) {};
\vertex [right=3.5cm of a,dot,label=below:{$\tau_5$}] (v5) {};
\vertex [right=4.2cm of a,dot,label=below:{$\tau_6$}] (v6) {};
\vertex [right=5cm of a,dot,label=below:{$\tau_7$}] (v7) {};
\vertex [right=6cm of a,dot,label=below:{$\tau_8$}] (v8) {};
\diagram* {
(a) -- [fermion,edge label'=\footnotesize \(\mathbf{k}_1\)] (v1) -- [fermion] (v2) -- [fermion] (v3) -- [fermion] (v4) -- [fermion] (v5) -- [fermion] (v6) -- [fermion] (v7) -- [fermion] (v8) -- [fermion,edge label'=\footnotesize \(\mathbf{k}_1\)] (b) 
(c) -- [boson, quarter left, edge label=\footnotesize \(\tilde{\mathbf{q}}_1\)] (v1),
(e) -- [boson, quarter left, edge label=\footnotesize \(\tilde{\mathbf{q}}_2\)] (v3),
(v6) -- [boson, quarter left, edge label=\footnotesize \(\tilde{\mathbf{q}}_1\)] (d),
(v8) -- [boson, quarter left, edge label=\footnotesize \(\tilde{\mathbf{q}}_2\)] (f),
(v2) -- [boson, half left, edge label=\footnotesize \(\mathbf{q}_1\)] (v7),
(v4) -- [boson, half left, edge label=\footnotesize \(\mathbf{q}_2\)] (v5),
};
\end{feynman}
\end{tikzpicture}
\caption{\label{fig:4thorder2Green} 8\textsuperscript{th}-order diagram for $G^{(2)}(\mathbf{k},\tau,\tilde{\mathbf{q}}_1,\tilde{\mathbf{q}}_2)$. Note that diagrams in the expansion of $G^{(2)}$ have two phonon propagators attached to the diagram end. The total polaron wave vector $\mathbf{k}=\mathbf{k}_1+\tilde{\mathbf{q}}_1+\tilde{\mathbf{q}}_2$ is conserved at the vertices.}
\end{figure}
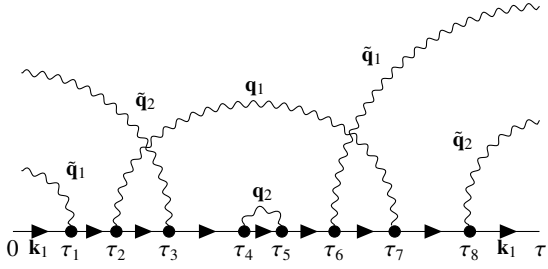

This allows us to write the Green's function as an infinite series over Feynman diagrams. Odd orders in the perturbation series evaluate to zero because phonon operators appear linear in the interaction term of the Hamiltonian (Eq.~\ref{eqn:interactionHamiltonian}). A typical diagram is presented in Fig.~\ref{fig:4thorder2Green}. It shows a 8\textsuperscript{th}-order diagram of $G^{(2)}(\mathbf{k},\tau,\tilde{\mathbf{q}}_1,\tilde{\mathbf{q}}_2)$. All diagrams of $G^{(N)}$ have $N$ external phonon propagators attached to the diagram end. The rules from Eq.~\ref{eqn:freeElectron} -~\ref{eqn:vertex} can be used to translate a diagram back into its functional form. Integration has to be performed over all internal phonon wave vectors $\{\mathbf{q}_i\}$ and over all times $\{\tau_i\}$ so that their chronological order is maintained, e.g. $0<\tau_1<\tau_2<\dots<\tau_8<\tau$ in Fig.~\ref{fig:4thorder2Green}. The total wave vector $\mathbf{k}$ is always conserved at interaction vertices. For example, the electron propagator between $\tau_1$ and $\tau_2$ in Fig.~\ref{fig:4thorder2Green} must have the wave vector $\mathbf{k}_2=\mathbf{k}_1+\tilde{\mathbf{q}}_1$ so that $\mathbf{k}=\mathbf{k}_2+\tilde{\mathbf{q}}_2$.

Expressing the Green's function in terms of Feynman diagrams doesn't solve the problem. It merely is a way to rewrite the expansion in a more accessible way. It is still necessary to sum the infinite series of integrals from Eq.~\ref{eqn:greenPertExpansion}.

\subsection{Diagrammatic Monte Carlo} \label{subsec:dmc}

In Ref.~\cite{Prokofev1998,Mishchenko2000,1063-7869-48-9-R02} it was shown how to use the DMC method to numerically calculate a function $Q(\{y\})$ which is given in a diagrammatic expansion of the form
\begin{equation}
Q(\{y\})=\sum_{n=0}^{\infty} \sum_{\xi_n} \idotsint \mathcal{D}_{n,\xi_n}(\{y\};x_1,\dots,x_n) \ dx_1\dots dx_n. \label{eqn:generalDMC}
\end{equation}
The overall idea behind the DMC method is to interpret $Q(\{y\})$ as a distribution function for the external variables $\{y\}$~\cite{Prokofev1998}. It then uses a Markov chain Monte Carlo (MCMC) procedure to simulate $Q(\{y\})$ by generating diagrams stochastically. This is achieved with a Metropolis-Hastings update scheme to accept or reject new diagrams in which the numerical values of $\mathcal{D}_{n,\xi_n}$ serve as statistical weights. The function $Q(\{y\})$ is obtained by collecting statistics for the external variables $\{y\}$, e.g. in the form of a histogram. At the heart of the DMC algorithm are updates that allow the Markov chain to explore the whole space of Feynman diagrams, i.e. the Markov chain has to be ergodic. It is therefore necessary to implement updates which change the order $n$, the topology $\xi_n$, external variables $\{y\}$ and internal variables $x_i$. Details on basic updating procedures and acceptance probabilities can be found in the Refs.~\cite{Prokofev1998,Mishchenko2000,VANHOUCKE201095,1063-7869-48-9-R02}.

\begin{figure}
\begin{alg}
\begin{algorithmic}
\STATE{\textbf{Input:} initial diagram $\mathcal{D}^{(0)}\gets (\{y^{(0)}\};x_1^{(0)},\dots,x_n^{(0)},n^{(0)},\xi_n^{(0)})$},
\STATE{\phantom{\textbf{Input:}} update procedures $\{U_1,\dots,U_k\}$,}
\STATE{\phantom{\textbf{Input:}} update probabilities $\{p(U_1),\dots,p(U_k)\}$;}
\STATE{\textbf{Output:} histogram of $Q(\{y\})$;}
\STATE{}
\STATE{initialize histogram[];}
\STATE{initialize diagram $\mathcal{D}_{cur}\gets \mathcal{D}^{(0)}$;}
\WHILE{not converged}
	\STATE{choose an update $U_i$ from $\{U_1,\dots,U_k\}$ with probability $p(U_i)$;}
	\STATE{propose a new diagram $\mathcal{D}_{new}\gets (\{y'\};x_1',\dots,x_{n'}',n',\xi_{n'}')$ ac- \\ \hspace*{7pt} cording to $U_i$;}
	\STATE{calculate acceptance ratio $R$;}
	\STATE{draw random uniform number $r$;}
	\IF{$R \geq r$}
		\STATE{accept the proposed diagram: $\mathcal{D}_{cur}\gets \mathcal{D}_{new}$;}
	\ELSE
		\STATE{reject the proposed diagram: $\mathcal{D}_{cur}\gets \mathcal{D}_{cur}$;}
	\ENDIF
	\STATE{histogram[$\{y\}$]$\gets$ histogram[$\{y\}$]$+1$;}
\ENDWHILE
\STATE{\textbf{return} histogram;}
\end{algorithmic}
\end{alg}
\caption{\label{fig:dmcWorkflow} General workflow of the DMC algorithm. The algorithm returns the histogram of the function $Q(\{y\})$.}
\end{figure}

A general workflow of a DMC application is sketched in Fig.~\ref{fig:dmcWorkflow}. Necessary requirements are a diagrammatic expansion of $Q(\{y\})$, updates $\{U_1,\dots,U_k\}$ and probabilities $\{p(U_1),\dots,p(U_k)\}$ with which the updates are chosen. The current diagram in each step is denoted by $\mathcal{D}_{cur}$ and characterized by its parameters values $\mathbf{z}=(\{y\};x_1,\dots,x_n,n,\xi_n)$. The proposed diagram is called $\mathcal{D}_{new}$ with new parameters $\mathbf{z}'=(\{y'\};x_1',\dots,x_{n'}',n',\xi_{n'}')$. At the beginning, an initial diagram $\mathcal{D}^{(0)}$, e.g. a free electron propagator, is defined and the grid for the histogram is generated. During each Monte Carlo step an update $U_i$ gets selected with probability $p(U_i)$. The update $U_i$ proposes a new diagram $\mathcal{D}_{new}$ by changing one or more of the current parameters of $\mathbf{z}$ to $\mathbf{z}'$. Then a Metropolis-Hastings accept/reject step is performed with the following acceptance ratio (detailed balance is assumed)
\begin{equation}
R = \frac{p(U_i^\dagger)\mathcal{D}_{new}P(\mathbf{z}'\to \mathbf{z})}{p(U_i)\mathcal{D}_{cur}P(\mathbf{z}\to \mathbf{z}')}, \label{eqn:acceptanceRatio}
\end{equation}     
where $p(U_i^\dagger)$ is the probability of selecting the inverse update $U_i^\dagger$ of $U_i$ and $P(\mathbf{z}\to \mathbf{z}')$ is an arbitrary probability density from which the new parameters $\mathbf{z}'$ are chosen. If $R\geq r$, where $r$ is a uniform random number, $\mathcal{D}_{new}$ is accepted otherwise rejected. Finally, the histogram at position $\{y\}$ is updated. These steps are repeated until convergence is achieved. Normalizing the resulting histogram leads to an estimation for $Q(\{y\})$.

\subsection{DMC for the Fr\"ohlich polaron}

With the general procedure of the DMC algorithm at hand, it is fairly easy to apply it to the Fr\"ohlich polaron. Comparing Eq.~\ref{eqn:greenPertExpansion} with~\ref{eqn:generalDMC} leads to the following identifications: 
\begin{enumerate}
 \item[(i)] $Q \leftrightarrow G^{(N)}$
 \item[(ii)] $\{y\} \leftrightarrow \{\mathbf{k},\tau,\{\tilde{\mathbf{q}}_i\}\}$
 \item[(iii)] $\{x_1,\dots,x_n\} \leftrightarrow \{\tau_1,\dots,\tau_n,\mathbf{q}_1,\dots,\mathbf{q}_k\}$
\end{enumerate}

The most straightforward way to obtain the lowest energy eigenvalues $E_{0}(k,\alpha)$ of the Fr\"ohlich Hamiltonian for a given $\mathbf{k}$ and $\alpha$ with the DMC method is to simulate $G^{(0)}(\mathbf{k},\tau)$ and fit an exponential function to its long imaginary time behaviour, as can be seen in Eq.~\ref{eqn:greenAsympt}. This was done in the original paper by Prokof'ev~\cite{Prokofev1998}.

Mishchenko \textit{et al.}~\cite{Mishchenko2000} provided some improvements to this method. They simulated all $G^{(N)}(\mathbf{k},\tau,\{\tilde{\mathbf{q}}_i\})$ up to some maximum value $N<N_{max}$ in a single run. It allowed them to introduce direct Monte Carlo estimators for the energy, effective mass, group velocity and Z-factors and to obtain results up to $\alpha=20$.

In the present paper, we follow the approach by Mishchenko using estimators for the energy $e_{est}(\mathcal{D})$ and inverse effective polaron mass $m_{est}(\mathcal{D})$ making the curve fitting procedure obsolete. A detailed exposition of the workflow can be found in Fig.~\ref{fig:dmcFrohWorkflow}. Values for the coupling constant $\alpha$ and the polaron wave vector $\mathbf{k}$ are defined as inputs before the simulation starts.  The parameter $\mu$ is used as part of a guiding function of the form $e^{\mu\tau}$ to improve the sampling in $\tau$-space. In practice this means that each diagram is multiplied by $e^{\mu\tau}$ or simply by changing the value of the free electron Green's function to
\begin{equation}
G_0(\mathbf{k},\tau_i-\tau_j,\mu)=e^{-(k^2/2-\mu)(\tau_i-\tau_j)}.
\end{equation}
For our calculations, we set $\mu$ slightly smaller than the true ground state energy, as recommended in Ref.~\cite{Prokofev1998}. We also have specified maximum values for the diagram length $\tau_{max}$, the order $n_{max}$ and for the number of phonon propagators attached to the diagram end $N_{max}$. The value $\tau_{min}$ is used as a cut off, in the sense that we only accumulate estimators if the current diagram length $\tau$ is greater than $\tau_{min}$. In our case, $\tau_{max}=50$ and $\tau_{min}=5$. Values for $n_{max}$ and $N_{max}$ are dependent on the coupling strength $\alpha$, $\tau_{max}$ and $\mu$ and should be chosen sufficiently higher than the average diagram order and average number of external phonons per diagram. The most important ingredients are the updates $U_i$. We implemented updates for adding and removing internal as well as external phonon propagators, changing the diagram length $\tau$, stretching the diagram as a whole, shifting a single vertex in imaginary time and swapping the phonon propagators of two adjacent vertices. All these updates and a derivation of the estimators are explained in detail in Ref.~\cite{Mishchenko2000}. We only changed the arbitrary proposal probability distribution $P(\mathbf{z}\to \mathbf{z}')$ for some of the updates (see Eq.~\ref{eqn:acceptanceRatio}). Updates are addressed with the same probability $p(U_i)=p(U_j)$.

\begin{figure}[t!]
\begin{alg}
\begin{algorithmic}
\STATE{\textbf{Input:} initial diagram $\mathcal{D}^{(0)}\gets (\mathbf{k},\tau^{(0)},\{\tilde{\mathbf{q}}_i^{(0)}\};\{\tau_i^{(0)}\},\{\mathbf{q}_i^{(0)}\},n^{(0)},\xi_n^{(0)})$},
\STATE{\phantom{\textbf{Input:}} update procedures $\{U_1,\dots,U_k\}$,}
\STATE{\phantom{\textbf{Input:}} update probabilities $\{p(U_1),\dots,p(U_k)\}$,}
\STATE{\phantom{\textbf{Input:}} values for: $\alpha$, $\mu$, $\mathbf{k}$,}
\STATE{\phantom{\textbf{Input:}} parameters: $\tau_{max}$, $\tau_{min}$, $n_{max}$, $N_{max}$;}
\STATE{\textbf{Output:} energy $E_{0}^{MC}(k,\alpha)$,}
\STATE{\phantom{\textbf{Output:}} inverse effective polaron mass $m_*^{MC}(\alpha)$;}
\STATE{}
\STATE{initialize diagram $\mathcal{D}_{cur}\gets \mathcal{D}^{(0)}$;}
\STATE{$E_{0}^{MC}\gets 0$, $m_*^{MC}\gets 0$;}
\STATE{$c\gets 0$;}
\WHILE{not converged}
	\STATE{choose an update $U_i$ from $\{U_1,\dots,U_k\}$ with probability $p(U_i)$;}
	\STATE{propose a new diagram $\mathcal{D}_{new}\gets (\mathbf{k},\tau',\{\tilde{\mathbf{q}}_i'\};\{\tau_j'\},\{\mathbf{q}_k'\},n',\xi_{n'}')$ ac- \\ \hspace*{7pt} cording to $U_i$;}
	\STATE{calculate acceptance ratio $R$;}
	\STATE{draw random uniform number $r$;}
	\IF{$R \geq r$}
		\STATE{accept the proposed diagram: $\mathcal{D}_{cur}\gets \mathcal{D}_{new}$;}
	\ELSE
		\STATE{reject the proposed diagram: $\mathcal{D}_{cur}\gets \mathcal{D}_{cur}$;}
	\ENDIF
	\IF{$\tau>\tau_{min}$}
		\STATE{$c\gets c+1$;}
		\STATE{$E_{0}^{MC}\gets E_{0}^{MC}+e_{est}(\mathcal{D}_{cur})$;}
		\STATE{$m_*^{MC}\gets m_*^{MC}+m_{est}(\mathcal{D}_{cur})$;}
	\ENDIF
\ENDWHILE
\STATE{\textbf{return} $E_{0}^{MC}/c$, $m_*^{MC}/c$;}
\end{algorithmic}
\end{alg}
\caption{\label{fig:dmcFrohWorkflow} Detailed workflow of the DMC algorithm as it was used in this paper. The algorithm returns estimates for the lowest eigenenergy $E_{0}(k,\alpha)$ and the inverse of the effective polaron mass $1/m_*(\alpha)$ for given $\mathbf{k}$ and $\alpha$ values.}
\end{figure}

\begin{figure}
\includegraphics{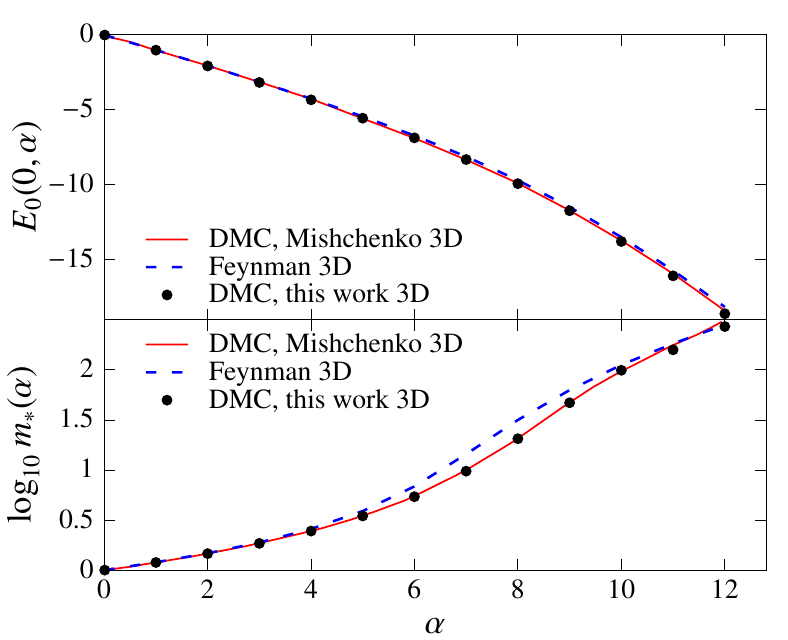}
 \caption{\label{fig:benchmark} Comparison of our results (circles) with previous DMC results by Mishchenko~\cite{Mishchenko2000} (continuous lines) and with results obtained with Feynman's approach 
 \cite{Rosenfelder2001} (dashed lines). The top graph shows the polaron ground state energy $E_{0}(0,\alpha)$ and the bottom graph the logarithm of the polaron effective mass $\log m_*(\alpha)$ as a function of $\alpha$.}
\end{figure}

The basic concept is the same as in the general DMC algorithm, except that we accumulate estimators instead of a histogram (cf. Fig.~\ref{fig:dmcWorkflow} and~\ref{fig:dmcFrohWorkflow}). We start from an initial diagram $\mathcal{D}^{(0)}$. The accumulators for the energy $E_{0}^{MC}$ and inverse effective mass $m_*^{MC}$ as well as the counter $c$, for the number of diagrams with $\tau>\tau_{min}$, are set to zero. In the main loop, an update $U_i$ is chosen with probability $p(U_i)$ and a new diagram $\mathcal{D}_{new}$ is proposed. It is accepted with probability $\min\{1,R\}$. After the accept/reject step, we check if the current diagram length is greater than $\tau_{min}$. If $\tau>\tau_{min}$, $c$ is increased by $1$ and the energy and inverse effective mass estimator for the current diagram $\mathcal{D}_{cur}$ are accumulated. 
The effective mass is calculated near $\mathbf{k}=0$ using the quadratic approximation:

\begin{equation}
m_*(\alpha)=\left[\frac{\partial^2 E_{0}(k,\alpha)}{\partial k^2}\right]_{k=0}^{-1}. \label{eqn:effMassDef}
\end{equation}  
The loop is repeated until the energy and inverse effective mass estimates have converged. The final estimates are obtained by dividing the accumulators by $c$.  

In Fig.~\ref{fig:benchmark}, we reproduced some of the results from Ref.~\cite{Mishchenko2000} to verify the correctness of our code. The top graph shows the polaron ground state energy and the bottom graph shows the logarithm of the effective mass as a function of $\alpha$. Our data are in very good agreement with Mishchenko's data which lets us assume that our code gives reliable DMC results. The figure also displays results obtained with Feynman's variational treatment~\cite{Rosenfelder2001}.

\section{Results and discussion} \label{sec:results}

In this section, we provide a more extensive discussion of the DMC results for the Fr\"ohlich polaron in 3D and 2D. We show and discuss polaron ground state energies, effective polaron masses and polaron dispersions for different coupling strengths and prove that DMC correctly accounts for the 3D$\rightarrow$2D scaling relations.  
All energies are given in units of $\hbar\omega_0$ and lengths in units of $\sqrt{\hbar/m\omega_0}$.

\subsection{Polaron ground state energy and effective mass}

\begin{figure}[b!]
\includegraphics{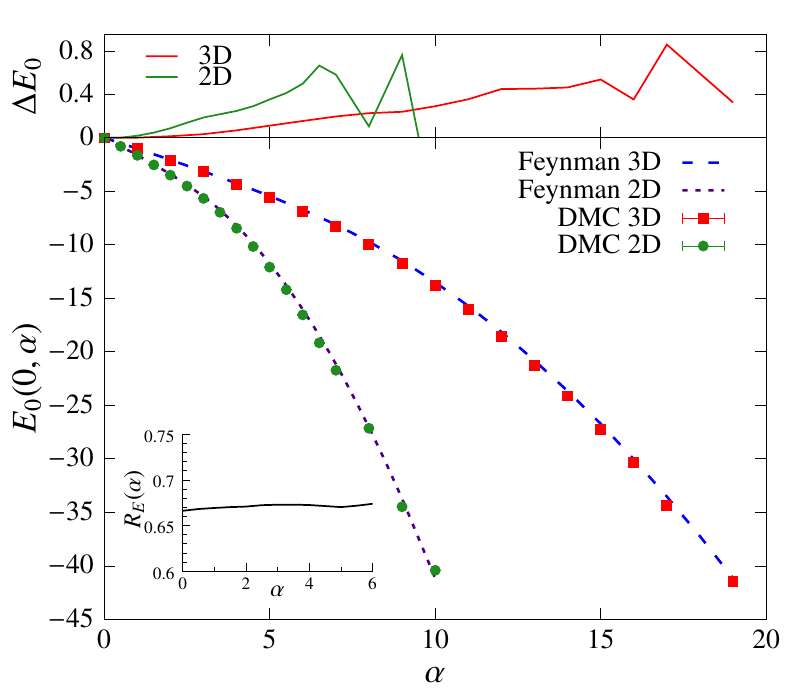}
 \caption{\label{fig:energy3d2d} Polaron energy $E_{0}(0,\alpha)$ as a function of the coupling constant $\alpha$. The modulus of the total wave vector is $k=0$. Results from the Feynman approach are shown as dashed lines. DMC results for 3D systems are depicted as squares and for 2D as circles. $\Delta E_0$ is the difference between Feynman and DMC results. The inset shows the scaling ratio $R_E(\alpha)=E_{0}^{2D}(0,\alpha)/E_{0}^{3D}(0, 3\pi\alpha/4)$ between our 2D and 3D DMC results.}
\end{figure}

\begin{figure}
\includegraphics{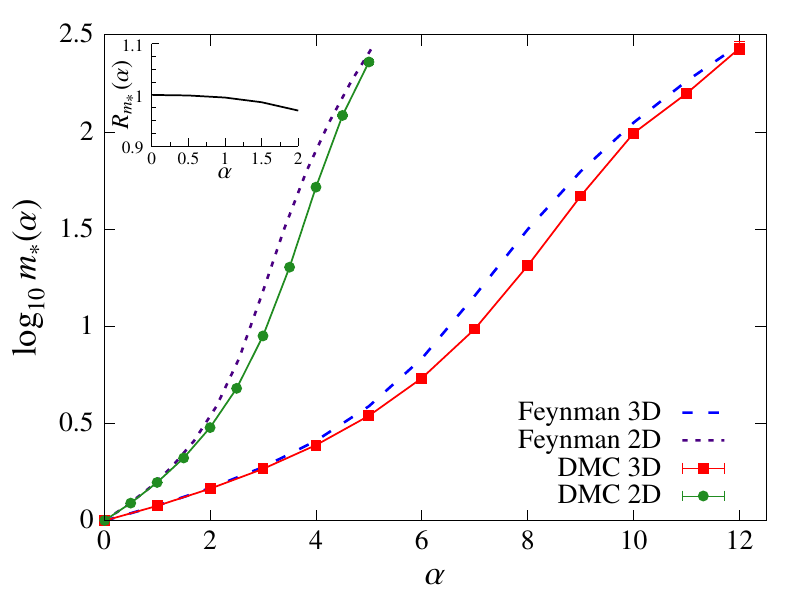}       
 \caption{\label{fig:mass3d2d} Logarithm of the polaron effective mass $m_*(\alpha)$ as a function of the coupling constant $\alpha$. Results from the Feynman approach are shown as dashed lines. DMC results for 3D systems are depicted as squares and for 2D as circles. The inset shows the scaling ratio $R_{m_*}(\alpha)=m^{2D}_{*}(\alpha)/m^{3D}_{*}\left(3\pi\alpha/4\right)$ between our 2D and 3D DMC results.}
\end{figure}

We first focus on our results for the polaron ground state energy $E_{0}(0,\alpha)$ (Fig.~\ref{fig:energy3d2d}), i.e. the minimum of the polaron energy band, and for the effective polaron mass $m_*(\alpha)$ (Fig.~\ref{fig:mass3d2d}) as a function of $\alpha$ for 3D and 2D systems. Both cases are compared to Feynman's approach~\cite{Rosenfelder2001} and with available DMC results in 3D~\cite{Mishchenko2000} (Fig.~\ref{fig:benchmark}). The corresponding numerical values are written in Table~\ref{tab:3d} (3D) and Table~\ref{tab:2d} (2D). 

Feynman results in 2D have been obtained from the 3D results via scaling relations~\cite{PhysRevB.36.4442,PhysRevB.31.3420,PhysRevB.37.933}. These scaling relations are exact for the Feynman polaron energy and Feynman polaron mass:
\begin{gather}
E_{0}^{2D}(0,\alpha) = \frac{2}{3}E_{0}^{3D}\left(0, 3\pi\alpha/4\right), \label{eqn:energyScaling} \\
\frac{m^{2D}_{*}(\alpha)}{m^{2D}} = \frac{m^{3D}_{*}\left(3\pi\alpha/4\right)}{m^{3D}}. \label{eqn:massScaling}
\end{gather}

For $\alpha=0$ the polaron does not form and therefore $E_{0}=0$ and $m_*(0)=m$. As expected, with increasing electron-phonon coupling the polaron energy $E_{0}(0,\alpha)$ decreases and the effective mass increases as a consequence of the progressive localization of the polaron band. This effect is stronger in 2D than in 3D and explains the steeper curves in 2D.

Overall, our DMC data agree very well with the Feynman results in the entire range of coupling strength, in particular for what concerns the polaron energy (Fig.~\ref{fig:energy3d2d}). The only sizeable deviation is observed for the effective mass in the intermediate coupling regime, for which Feynman's approach gives considerably higher values than the DMC (Fig.~\ref{fig:mass3d2d}).
Both the DMC results and the variational results obey the scaling laws (\ref{eqn:energyScaling}) and (\ref{eqn:massScaling}). This can be seen in the insets of Figs.~\ref{fig:energy3d2d} and~\ref{fig:mass3d2d} where we show the ratios $R_E(\alpha)=E_{0}^{2D}(0,\alpha)/E_{0}^{3D}(0, 3\pi\alpha/4)$ and $R_{m_*}(\alpha)=m^{2D}_{*}(\alpha)/m^{3D}_{*}\left(3\pi\alpha/4\right)$ between our DMC results in 2D and 3D. 
However, the uncertainty in the Monte Carlo calculations of $m^{2D}_{*}$ for $\alpha>2$ worsens the 
stability of the scaling relation of the effective mass at large $\alpha$.
The reason for this low performance is that the effective mass estimator actually calculates the inverse of the effective mass rather than the effective mass itself~\cite{Mishchenko2000}. Since the polaron mass grows very fast with increasing coupling, its inverse becomes very small, which unavoidably worsens the accuracy of the results.

\begin{table}
\caption{\label{tab:3d} Ground state energies $E_{0}(0,\alpha)$ and effective masses $m_*(\alpha)$ in 3D from the DMC and Feynman method~\cite{Rosenfelder2001}. Values in brackets stand for the uncertainty in the DMC simulation, e.g $-1.01662(47)$ has a sample standard error of $4.7\times10^{-4}$.}
\begin{ruledtabular}
\begin{tabular}{ccccc}
\multicolumn{1}{c}{$\alpha$} & \multicolumn{1}{c}{$E_{0}$ DMC} & \multicolumn{1}{c}{$E_{0}$ Feynman} & \multicolumn{1}{c}{$m_*$ DMC} & \multicolumn{1}{c}{$m_*$ Feynman} \\ 
\hline 
1 & -1.01662(47) &	-1.0130308 & 1.19396(2) & 1.1955147 \\
2 &	-2.06957(84) & -2.0553559 & 1.46166(7) & 1.4718919 \\
3 & -3.16829(136) & -3.1333335 & 1.85047(13) & 1.8889540 \\
4 & -4.32490(211) & -4.2564809 & 2.45196(57) & 2.5793104 \\
5 & -5.55297(296) & -5.4401445 & 3.47194(180) & 3.8856197 \\
6 & -6.86647(287) & -6.7108710 & 5.41952(625) &	6.8383564 \\
7 & -8.31039(309) & -8.1126875 & 9.7130(268) & 14.394070 \\
8 & -9.92206(606) & -9.6953709 & 20.55(14) & 31.569255 \\
9 & -11.72535(701) & -11.485786 & 46.90(78) & 62.751527 \\
10 & -13.7820(136) & -13.490437 & 98.8(3.3) & 111.81603 \\
11 & -16.0660(127) & -15.709808 & 158.2(4.6) & 183.12497 \\
12 & -18.5943(240) & -18.143395 & 270.1(20.0) & 281.62189 \\
13 & -21.2434(249) & -20.790681	& / & 412.78190 \\
14 & -24.1151(369) & -23.651278	& / & 582.58390 \\
15 & -27.2629(359) & -26.724904	& / & 797.49838 \\
\end{tabular}
\end{ruledtabular}
\end{table}

\begin{table}
\caption{\label{tab:2d} Ground state energies $E_{0}(0,\alpha)$ and effective masses $m_*(\alpha)$ in 2D from the DMC and Feynman method~\cite{Rosenfelder2001}. Values in brackets stand for the uncertainty in the DMC simulation, e.g $-1.64348(23)$ has a sample standard error of $2.3\times10^{-4}$.}
\begin{ruledtabular}
\begin{tabular}{ccccc}
$\alpha$ & $E_{0}$ DQMC & $E_{0}$ Feynman & $m_*$ DQMC & $m_*$ Feynman \\ 
\hline 
1 & -1.64348(23) & -1.62321 & 1.57437(8) & 1.59966 \\
2 &	-3.48333(62) & -3.39482 & 3.01609(21) & 3.40982 \\
3 & -5.66337(46) & -5.47667 & 8.94191(730) & 15.2085 \\
4 & -8.45543(149) & -8.20738 & 52.108(341) & 81.1684 \\
5 & -12.08288(610) & -11.7281 & 229.3(7.8) & 257.452 \\
6 & -16.5403(269) & -16.0402 & 601.9(46.0) & 609.244 \\
7 & -21.7231(566) & -21.1408 & / & / \\
8 & -27.1346(802) & -27.0283 & / & / \\
9 & -34.4669(370) & -33.7021 & / & / \\
10 & -40.4139(379) & -41.1602 & / & / \\
\end{tabular}
\end{ruledtabular}
\end{table}

\begin{table*}
\caption{\label{tab:expansionCoeff} Exactly known (exact) vs. calculated (calc.) expansion coefficients of $E_{0}(0,\alpha)$ for the weak- and strong coupling limit. The coefficients were obtained using different 
ranges of $\alpha$ in 2D and 3D. In 2D, we have included $\alpha<0.2$ for computing $q_1$ and $q_2$ 
and $4\le\alpha<9$ for $\gamma$. The corresponding 3D ranges are $\alpha<0.85$ ($q_1$ and $q_2$)
and $9\le\alpha<18$ ($\gamma$).}
\begin{ruledtabular}
\begin{tabular}{c|cccccc}
 & $q_1$ exact & $q_1$ calc. & $q_2$ exact & $q_2$ calc. & $\gamma$ exact & $\gamma$ calc. \\
\hline
3D & 1.0 & 0.9999 $\pm$ 3.8$\times 10^{-4}$  & 0.01592 & 0.01588 $\pm$ 9.1$\times 10^{-4}$  & 0.1085 & 0.10805 $\pm$ 7.7$\times 10^{-4}$ \\
2D & 1.5708 & 1.57084 $\pm$ 1.7$\times 10^{-4}$ & 0.06397 & 0.06483 $\pm$ 2.8$\times 10^{-3}$ & 0.4047 & 0.40236 $\pm$ 3.8$\times 10^{-3}$ \\
\end{tabular}
\end{ruledtabular}
\end{table*}

To test the accuracy of our calculations, we have also retrieved values for the exactly known 
weak-coupling coefficients $q_1$ and $q_2$
\begin{equation}
E_{0}(0, \alpha) = -q_1\alpha - q_2\alpha^2 + \mathcal{O}(\alpha^3) \label{eqn:weakExpansion}
\end{equation}
and the strong-coupling coefficient $\gamma$
\begin{equation}
\lim_{\alpha \to \infty} E_{0}(0, \alpha)/\alpha^2 = -\gamma. \label{eqn:strongExpansion}
\end{equation} 

The exact~\cite{PhysRevB.31.3420,Gerlach2003} and DMC values for these coefficients, listed in 
Table~\ref{tab:expansionCoeff}, are in very good agreement.
However, a word of caution is needed here: the coefficients are obtained with a simple curve fitting procedure and the final numerical values are highly sensitive to the range of $\alpha$ values included in the fitting process. We have computed $q_1$ and $q_2$ using $\alpha<0.85$ and $\alpha<0.2$, in 3D and 2D respectively, whereas for $\gamma$ we have included values in the range $9\le\alpha<18$ (3D) and  $4\le\alpha<9$ (2D). 

Gerlach, Kalina and Smondyrev~\cite{Gerlach2003} correctly point out that the (3D) second order perturbative result $q_2=0.0126$ obtained by Mishchenko using DMC~\cite{Mishchenko2000} deviates from R\"oseler's~\cite{Roseler1968} exact result $q_2=0.01592...$, but we surmise that they incorrectly concluded that the DMC results $E_0$(0,$\alpha$) are incompatible with R\"oseler's results.
Here, we resolve this issue by providing the calculated DMC values explicitly, showing that there is no discrepancy. Both for the 3D and the 2D case, it can be seen in Table~\ref{tab:expansionCoeff} that the DMC technique yields accurate estimates for $q_2$, as well as for the other analytically known expansion coefficients $q_1$ and $\gamma$.

\subsection{Polaron dispersion}

\begin{figure*}
\includegraphics{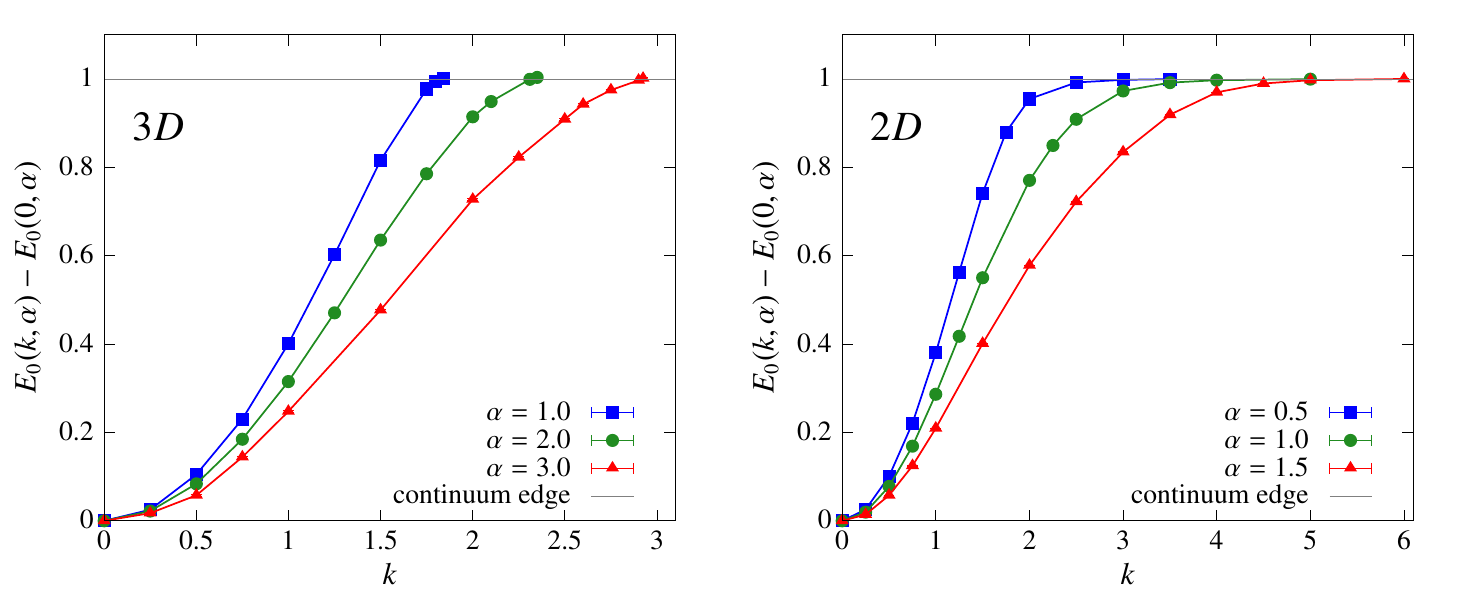}
 \caption{\label{fig:multi_disp} Polaron energy $E_{0}(k,\alpha)-E_{0}(0,\alpha)$ as a function of the modulus of the total wave vector $k$ in 3D (left, for coupling constants $\alpha=1.0, 2.0$ and 3.0) and 2D (right, $\alpha=0.5, 1.0$ and 1.5). The continuum edge is shown at $E_c(k)=1$.}
\end{figure*}

In Fig.~\ref{fig:multi_disp}, we display some dispersion curves in 3D and 2D for selected values of
$\alpha$. The results have been shifted so that the ground state energy at $k=0$ is $E_{0}(0,\alpha)=0$.
This makes a comparison between different $\alpha$ values easier. 
As expected, $E_{0}(k,\alpha)$ increases monotonically as a function of $k$ and becomes more flat with
increasing coupling.  This reflects the tendency to form more localized bands as the electron-phonon
coupling strength becomes stronger, an effect that is more intense in the more-localized 2D limit, 
where the dispersion curves bend over more sharply. 
Clearly, this behavior correlates with the polaron effective mass since it is defined as the inverse of the curvature of the energy band at $k=0$ (see Fig.~\ref{fig:mass3d2d}).

For large $k$, the energy curve approaches the so called "continuum edge"
$E_c(\alpha)$ defined as the energy value:
\begin{equation}
E_c(\alpha)=E_{0}(0,\alpha)+\hbar\omega_0=E_{0}(0,\alpha)+1,
\end{equation}
i.e. the energy value which is one phonon excitation quantum or unity (in our units) above the ground state energy. An important difference between the 3D and 2D case is that in 3D the dispersion curve crosses the continuum edge at a finite critical wave vector length $k_c(\alpha)$. Instead, in 2D, it has been proven that this edge constitutes an asymptote and is approximated from below as $k\to \infty$~\cite{Gerlach2003,PhysRevB.77.174303,PhysRevB.60.10886}.

\begin{figure*}
\includegraphics{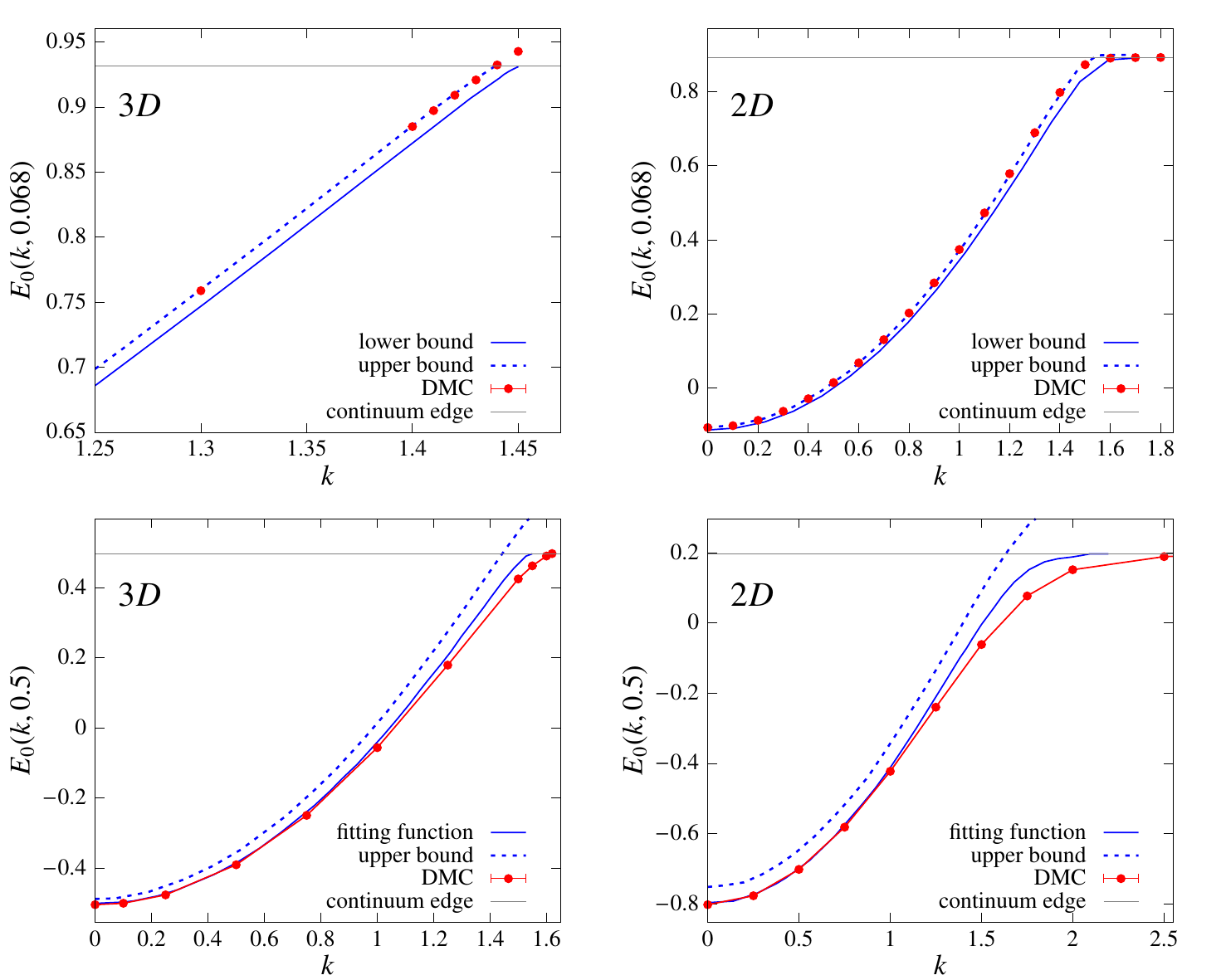}
 \caption{\label{fig:multi_gerlach} Polaron energy $E_{0}(k,\alpha)$ in 3D (left) and 2D (right) as a function of the modulus of the total wave vector $k$ for coupling constant $\alpha=0.068$ (top row) and $\alpha=0.5$ (bottom row). Lower and upper bounds, and a fitting function to the dispersion are taken 
 from Ref.~\cite{PhysRevB.77.174303}.}
\end{figure*}

For small $\alpha$, there exist rigorous upper and lower bounds for the polaron 
dispersion~\cite{PhysRevB.77.174303} that restrict this dispersion to a narrow domain. In the top row of
Fig.~\ref{fig:multi_gerlach}, the DMC results are shown together with these bounds for $\alpha=0.068$,
the value of the coupling strength for GaAs. 
Our results 
lie in between the bounds, close to the upper bound, both in 3D (upper left panel of 
Fig.~\ref{fig:multi_gerlach}) and 2D (upper right panel). The strict lower bound only exists for small values of the coupling strength: $\alpha=0.5$ already lies outside the range where this lower bound can be found.

Gerlach and Smondyrev~\cite{PhysRevB.77.174303} propose a fitting function for the dispersion. 
This fit is based on a re-scaling of the upper bound formula, to obtain the correct gap between 
bottom of the band and the continuum edge, while maintaining the effective mass. 
As shown in the lower left panel of Fig.~\ref{fig:multi_gerlach}, the DMC results for the 3D case 
for $\alpha=0.5$ lie below both the variational upper bound and the Gerlach-Smondyrev dispersion. 
The same conclusion  can be drawn for the 2D case, shown in the lower right panel of Fig.~\ref{fig:multi_gerlach}.

\begin{table}
\caption{\label{tab:kc} Critical wave vectors $k_c(\alpha)$ for coupling constants 
$\alpha=0.068$, $\alpha=0.5$ and $\alpha=1.0$. Listed are results from our DMC calculations, from
Eq.~\ref{eqn:criticalLength} which is valid up to first order in $\alpha$, as well as from the fitting
function from Ref.~\cite{PhysRevB.77.174303}.}
\begin{ruledtabular}
\begin{tabular}{l|ccc}
 & $\alpha=0.068$ & $\alpha=0.5$ & $\alpha=1.0$ \\
\hline
DMC, this work & 1.440 & 1.615 & 1.833 \\
Result to order $\alpha$, Eq.~(\ref{eqn:criticalLength}) & 1.442 & 1.616 & 1.818 \\
Gerlach and Smondyrev, Ref.~\cite{PhysRevB.77.174303}  & 1.442 & 1.570 & 1.697 \\
\end{tabular}
\end{ruledtabular}
\end{table}

We now focus on the 3D case, in which the dispersion reaches the continuum edge at a given $k_c$. Up to lowest order in $\alpha$, 
\begin{equation}
  k_c(\alpha)=\sqrt{2} + \left(\frac{\pi}{2} - 1\right)\frac{\alpha}{\sqrt{2}} + \mathcal{O}(\alpha^2). 
\label{eqn:criticalLength}
\end{equation}
In Table~\ref{tab:kc}, we compare for several $\alpha$ values the critical wavenumber obtained (i) with 
DMC, (ii) with the first order approximation, Eq.~\ref{eqn:criticalLength}, and 
(iii) using the Gerlach-Smondyrev dispersion. 
At small coupling strength $\alpha=0.068$, all three approaches yield the same result. 
However, as $\alpha$ is increased slightly (remaining in the regime 
where the lowest order approximation can be expected to be valid), the result obtained from the 
Gerlach-Smondyrev dispersion drops below the value found by the other two approaches. The value of 
$k_c$ in the Gerlach-Smondyrev approach is 3\% resp.~8\% smaller than the DMC result for 
$\alpha=0.5$ and 1.

Previously~\cite{PhysRevB.77.174303}, this discrepancy was blamed on the fact that the 
DMC method supposedly failed to reproduce even the known $q_2$ parameter (the coefficient of $\alpha^2$), 
whereas the fitting function is claimed to be good up to order $\alpha^3$. However, as we have shown in 
the previous subsection, this explanation cannot hold since contrary to what was believed earlier, 
the DMC does reproduce the $q_2$ value with high accuracy. 
The Gerlach-Smondyrev dispersion is not the result of variational minimization, 
nor is it a rigorous lower bound: rather it is an ad hoc proposal that rescales the best 
variational upper bound to give the correct known limits. Keeping in mind that the DMC 
calculation takes many phonons into account (i.e. goes well beyond order $\alpha$ in the diagrams), 
we can conclude that the DMC results indicate that this fitting procedure is not appropriate
for $\alpha \ge 0.5$.

\section{Summary and Conclusion} \label{sec:conclusion}

The Diagrammatic Monte Carlo is a powerful method which has proven to work in many applications for many different systems~\cite{PhysRevB.87.115133,PhysRevLett.87.186402,PhysRevLett.113.166402,PhysRevLett.91.236401,PhysRevB.89.085119,1367-2630-17-3-033023}. For this paper, we have implemented a DMC code based on the Refs.~\cite{Prokofev1998,Mishchenko2000} and applied it to the solution of the large polaron Fr\"ohlich Hamiltonian in 3D and 2D. We benchmarked our code with existing DMC results for the 3D case to verify its correctness and then computed polaron ground state energies, effective polaron masses and polaron dispersion curves in 2D and 3D. 

In summary, our data confirm that the effect of electron-phonon coupling is enhanced in 2D compared to 3D, 
and this is reflected in all computed physical quantities. Concerning the ground state energies, the DMC 
results are in very good agreement with those obtained by Feynman's approach~\cite{Rosenfelder2001} and 
we have demonstrated that they obey the scaling relations between 3D and 2D~\cite{PhysRevB.36.4442}.
The reliability of the DMC procedure is further corroborated by the calculations of the coefficients used 
for the weak- and strong-coupling regime, which are almost identical to the exactly known values. 
This refutes a claim~\cite{Gerlach2003} that the DMC technique is not able to correctly obtain the $q_2$ coefficients. Regarding the effective polaron mass, the DMC performance becomes slightly less satisfactory at stronger coupling. This inaccuracy should be traced back to the numerical errors involved in the calculation of the inverse of the effective mass.
Alternative definitions of the polaron effective mass have been proposed in literature, which could be possibly tested in future work to assess and compare the performance of DMC and path-integrals approaches~\cite{PhysRevB.37.3085,PhysRevB.57.8739}.

One of the most interesting outcomes of the present study are the polaron dispersion curves. 
The DMC calculations reproduce very well the different behaviour seen in 2D and 3D: in 2D the energy curve 
approaches the continuum edge asymptotically from below, whereas in 3D it reaches the continuum edge at 
a finite critical $k_c$.
For small $\alpha$ (=0.068, a realistic value for a material like GaAs), the DMC 
dispersion as well as the $k_c$ are in very good agreement with the known lower and upper limits derived
from the variational approach of Gerlach and Smondyrev~\cite{PhysRevB.77.174303}. 
For larger $\alpha$ ($\alpha$= 0.5, 1.0), the DMC data agree well with the first order 
expansion results, but deviate from the values based on a proposed fitting function for 
the dispersion. 
While the DMC technique cannot validate the fitting procedure proposed by Gerlach and Smondyrev 
for $\alpha \ge 0.5$, it does suggest that up to $\alpha \approx 1$ the first order expansion 
result of Eq.~\ref{eqn:criticalLength} already provides an accurate estimate of $k_c$. 

\section*{Acknowledgements} \label{sec:acknowledgements}

This work was supported by the joint FWO-FWF project POLOX (Grant No. I 2460-N36).
Supercomputing time on the Vienna Scientific cluster (VSC) is gratefully acknowledged.

\bibliographystyle{apsrev4-1}
\bibliography{reference} 

\end{document}